# Speech recognition assisted by large language models to command software orally. Application to an augmented/virtual reality web app for immersive molecular graphics.

*Fabio Cortés Rodríguez and Luciano Abriata*

*EPFL Lausanne, Switzerland*

**Summary**

This project successfully developed, evaluated and integrated a Voice User Interface (VUI) into a web application that we are developing for immersive molecular graphics. Said app provides augmented/virtual reality (AR/VR) environments where users manipulate molecules with their hands, but this means the hands can't be used to control the app through a regular mouse- and keyboard-based GUI. The speech-based VUI system developed here alleviates this problem, making it easy to control the app via natural spoken (or typed) commands. To achieve this VUI we evaluated two distinct Automated Speech Recognition (ASR) systems: Chrome's native Speech API and OpenAI's Whisper v3. While Whisper offered broader browser compatibility, its tendency to "hallucinate" with specialized scientific jargon proved very problematic. Consequently, we selected Chrome's ASR for its stability, speed, and reliability. For translating transcribed speech into software commands, we tested two Large Language Model (LLM)-driven approaches: either generating executable code, or calling predefined functions. The function call method, powered by OpenAI's GPT-4o-mini, was ultimately adopted due to its superior safety, efficiency, and reliability over the more complex and error-prone code-generation approach. The resulting VUI is then based on an integration of Chrome's ASR with our LLM-based function-calling module, enabling users to command the application using natural language as shown in a video linked inside this report. We provide links to live examples demonstrating all the intermediate components, and details on how we crafter the LLM's prompt in order to teach it the function calls as well as ways to clean up the transcribed speech and to explain itself while generating function calls. For best demonstration of the final system, we provide a video example.

**Introduction**

Recent work has explored adaptive edge-cloud orchestration of ASR paired with LLMs to convert voice input into actionable commands for downstream systems.[1] In industrial settings, LLM-enhanced Voice User Interface (VUI, as opposed to Graphical User Interface, GUI) have been shown to mitigate limitations of traditional interfaces, emphasizing flexibility and natural language grounding in control tasks.[2,3] The core hypothesis motivating this work is that, likewise, VUIs should be of great use in augmented and virtual reality scenarios.

This document reports the development and evaluation of a system designed to integrate speech recognition (ASR) and natural language processing (NLP) capabilities via large language models (LLMs) into software for molecular graphics and modeling using augmented and virtual reality (AR/VR). The study took place within our overarching goal of enhancing user interaction with software by enabling voice-based commands in our AR/VR tools[4–7] while maintaining its flexibility and functionality across different devices and platforms. More specifically, in this work we explored how to create a VUI that can be plugged into a specific well-contained web app that we are developing for molecular visualization and modeling in multiuser AR and VR. Such web app is very well-suited as a container of this project because users inside AR/VR sessions typically have their hands busy engaging with the virtual objects, therefore a speech-based system for interaction with the device and software could potentially be extremely useful.

This report outlines the methodologies employed throughout the work, the performance of the different ASR and NLP components evaluated, their integration with other components of our AR/VR web app for molecular graphics and modeling, and its potential broader impact; while also addressing challenges and areas for possible improvements. Throughout we provide links to web pages that implement our partial tests as well as screenshots documenting example outputs from the ASR, NLP, and ASR-NLP pipelines.

### Section 1: Automated Speech Recognition (ASR) (or Text-To-Speech, TTS) systems

We tested two distinct ASR systems during this project: Chrome's native speech recognition system as available in the Speech API, which this browser is among the few to support; and Whisper v3, OpenAI's flagship model for ASR built on audio token recognition coupled to language processing. Chrome's Speech API is natively built into the browser itself in its versions for smartphones, tablets and computers, making it extremely easy to use and write code for. In addition, it can be used at zero cost. Whisper v3 is open source which means it can in principle be installed and deployed for free too, but in a real scenario its deployment in a web-accessible form is not trivial and certainly not achievable for free. To avoid the work and costs associated to this, we opted to use Whisper v3 via an API provided by a company called Gladia, which provided us with free credits to carry out various tests.

The facts that Chrome's ASR system works natively within the browser and that not many other browsers support the Speech API, generates a browser dependency that limits its application especially in immersive environments where Chrome just isn't supported, and although there are Chromium-based web browsers for AR/VR headsets, they do not support the Speech API. On the other hand, Whisper v3 accessed through Gladia's API abstracts the complexities of running the model, providing an infrastructure that ensures responsive and efficient performance at low (but non-zero) cost. Importantly, for our work here, Gladia offered us not only free credits for testing but also sufficient pieces of JavaScript code that facilitated integration into web apps through sockets.

In our own tests, the Whisper solution demonstrated the expected compatibility across all tested browsers, including Chrome, Brave, Firefox, and Safari, on laptops and smartphones. Crucially, it also worked with the Meta Quest Browser on AR/VR headsets like the Meta Quest 2 and 3, enabling voice interactions directly within immersive environments. This capability was a significant advantage for our application, as the molecular graphics tool is intended to operate within AR/VR settings where traditional interfaces may not be practical but the Meta Quest Browser does not support the Speech API.

Finally, it is important to acknowledge that beyond recognition accuracy, recent work has evaluated the energy and environmental cost of different ASR systems, a practical consideration for deployed voice UIs.[8]

*Example code and web pages for the first testing of both ASR systems*

The most basic implementations of Chrome and Whisper ASR systems are available at the following links:

- http://lucianoabriata.altervista.org/VUIsProject/basic-deploy---Chrome-STT.html

- http://lucianoabriata.altervista.org/VUIsProject/basic-deploy---Whisper-STT.html

Note: For the Whisper app a Gladia API key is required.

The following two examples incorporate customized grammar expansion, tested as a way to overcome the limitations of ASR systems when they must detect and transcribe highly specialized jargon:

- https://lucianoabriata.altervista.org/VUIsProject/grammar-expansion---Chrome-STT.html (works in Chrome only)
- https://lucianoabriata.altervista.org/VUIsProject/grammar-expansion---Whisper-STT.html

*Performance observations*

In terms of performance, Chrome's ASR system demonstrated certain reliability issues. After extended periods of listening, it occasionally stopped functioning. This problem could however be mitigated quite reliably by implementing a programmatic restart of the ASR system at key points of the web app's code. Other than this negative point, Chrome's ASR was stable and slightly faster than Whisper.

Whisper, while more flexible in terms of device and browser compatibility, presented challenges due to hallucination, that is the continued output of tokens unrelated to the actual input. This issue was particularly pronounced when the system was prompted with grammar or specialized jargon, a substantial problem because as described in the original project proposal we intended to use grammar injection in order to improve recognition of words that are relevant to the molecular sciences but are not frequent in everyday conversations. For instance, Whisper sometimes generated nonsensical or non-alphanumeric outputs when transcribing domain-specific terms, making downstream cleanup and interpretation difficult. Although Whisper performed relatively well for general speech in English, Spanish, French, and Italian, its struggles with specialized vocabulary may be attributed to the limited representation of such jargon in its training data. Eventually, we found that this limitation hindered its applicability to the highly specialized jargon of molecular modeling.

*Decision for one ASR system*

Ultimately, then, we opted for Chrome's ASR system due to its greater stability, lower cost, and faster performance. While this choice came at the expense of compatibility with non-Chrome browsers, especially in AR/VR headsets, the trade-off was deemed acceptable given the immediate needs of the project and the system's extremely simple implementation.

**Section 2: Speech-to-command casting**

*Approaches tested*

As outlined in the original proposal, we coded and tested two approaches to translate user commands into actions via an LLM, within the context of our growing molecular graphics tool. The first approach involved calling predefined functions, while the second entailed generating executable code based on the transcriptions. The latter is expected to be much more flexible if working properly, but at the expense of safety concerns because we cannot rule out that malicious code could be created and executed. Moreover, we eventually found other problems, as described below.

In turn, the first approach (calling of predefined functions) lacks flexibility unless a sufficient number of functions are hard-coded and predefined; but to its advantage, this approach runs faster, consumes less LLM tokens, and entails no safety issues because no code is executed—predefined functions that already exist in the code are simply detected in the LLM's output and then parsed and called. This approach relied on a carefully crafted prompt for OpenAI's GPT-4o-mini, the LLM of choice in our project as it is fast and inexpensive. Over several iterations, the prompt was refined to accurately map transcriptions to specific functions, ensuring consistent performance. This approach proved highly effective, even correcting transcription errors in some cases. By feeding well-defined examples into the prompt, the system was able to interpret user commands reliably and execute the corresponding actions. Additionally, we observed that providing comments that explain why and how each function is called in the prompt's examples, resulted in better generations.

The second approach, i.e. the one based in generating executable scripts, demonstrated significant potential for flexibility. It was in fact surprising to see that with sufficient time and tokens, the approach that generates and executed code within the tool was very powerful and promising, provided the safety issues could be cleared out. However, testing the approach highlighted some challenges beyond those expected regarding security risks. First, the much longer prompts (as they need to demonstrate not just function calls but actually several lines of code for each expected task it should accomplish) and output token requirements increased complexity, slowed execution, and raised costs. For instance, the number of tokens required for this approach was five to ten times greater than for predefined function calls (because often several lines of code had to be generated, vs. just a few function calls in the first approach). Second, we observed that when the prompt got too long the probability of errors increased substantially, and in certain cases it even happened that the LLM's output got truncated, all of which would have fatal consequences on program execution.

*Final implementation*

Considering the above challenges, we adopted the predefined function approach as the primary method for speech-to-command translation. This method allowed us to balance reliability and efficiency while maintaining large flexibility to expand the tool's functionality. For every new feature added to the tool, corresponding examples were incorporated into the prompt, ensuring that the system could handle the expanded functionality seamlessly.

Here's an excerpt put together from the core prompt we have crafted (as of the time of submission of this report) to teach the LLM how to call functions inside our app, followed by the first few examples of input-output pairs:

```
{ role: "system", content: "You receive texts from speech recognition and convert them to
commands by following the examples. If the texts you receive are in a language other than
English, you just translate to English and then execute. If you don't understand the request,
you run the command didntUnderstand(). The only functions you can output are
countAtoms(),acknowledge(), sayTime(), sayDate(), zoomIn(), zoomOut(), changeTemperature(),
changeUpdateRate(), setTemperature(), startSimulation(), stopSimulation(), writePDB(),
select(), changeUpdateRate(), speakUp(), spacefill, sticks, color, and didntUnderstand(). For
requests without any close examples, you do smart conversions, for example 'cartoon' refers to
using a backbone selection and then applying 'spacefill 1; sticks 2'. Make sure you don't use
things that don't exist, such as color byref. For requests without any close examples, you do
smart conversions, for example 'cartoon' refers to atoms that make up a backbone, such as N, C
and CA atoms in proteins or P atoms in nucleic acids. Note that some words might be wrong in
the input provided due to errors upon speech recognition, for example 'some' might need to be
replaced by 'zoom', 'so mean' could be 'zoom in', 'change' might actually be 'chain' or 'chains',
'handball' would actually be 'HandMol', 'mini mice' would be 'minimize', etc. Finally, make sure
you output comments explaining why each function is called, as shown in the examples."

{ role: "user", content: "Tell me the number of atoms" },

{ role: "assistant", content: " //Use countAtoms()\ncountAtoms();" },

{ role: "user", content: "Increase temperature" },

{ role: "assistant", content: " //There's a specific command to change temperature by a value,
and another to set the temperature to a given value. This request asks for an increase, so we
use changeTemperature() passing a positive number as argument\nchangeTemperature(+30);" },
```

```
{ role: "user", content: " Show all positively charged residues as spheres." },

{ role: "assistant", content: "//Simply select the positively charged amino acids, that is ARG and LYS, and show them as spheres of radius 3 with 'spacefill 3'\nselect('resname ARG LYS'); spacefill 3;" },

{ role: "user", content: "Stop simulation" },

{ role: "assistant", content: "//User asked to stop (pause) the simulation\nstopSimulation();"},

{ role: "user", content: "Write the coordinates" },

{ role: "assistant", content: "//Call the writePDB() function\nwritePDB();" },

{ role: "user", content: "Highlight residue number 1" },

{ role: "assistant", content: "//To highlight an atom we make its sphere bigger, so 'spacefill 3' goes here\nselect('resid 1'); spacefill 3; sticks 1;" },

{ role: "user", content: "No wait but show it in red" },

{ role: "assistant", content: "//We first re-select the atoms, and then apply the color\nselect('resid 1'); color red;" },

{ role: "user", content: "Color chain C by charge" },

{ role: "assistant", content: "//Select chain C specifically and then apply the standard color charges for amino acid charges\nselect('chain C'); color white; select('resname ASP GLU and chain C'); color red; select('resname ARG LYS and chain C'); color blue;",

{ role: "user", content: "Make spheres bigger" },

{ role: "assistant", content: "//As no specific atoms or molecules are indicated, we select all and increase the radius\nselect('all'); spacefill 3;" },

{ role: "user", content: "Much bigger" },

{ role: "assistant", content: "select('all'); spacefill 6;" },

},
```
…

A few relevant notes at this point. First, the above excerpt only shows representative examples of user/assistant pairs. The full prompt as of the date of submission of this report is around 300 lines of code, for around 9,000 tokens. Second, the selected examples show some of the features we incorporated, both at the app-specific level (controlling simulations, changing views and colors, requesting information about the system, etc.) and a general VUI level, for example the possibility of better understanding requests based on context (see for example "*Make spheres bigger*" followed by "*Much bigger*" towards the end of the prompt). Third, note that many of the example outputs include a comment in JS format i.e. preceded by two slashes (//); we found that this makes the outputs better, probably as they help the LLM to "think" about the user's request more carefully.

We leave the web pages built to test code generation and function calling with the Chrome and Whisper ASR systems (API keys are required):

- https://lucianoabriata.altervista.org/VUIsProject/request-to-code---Chrome-STT.html (works in Chrome only)
- https://lucianoabriata.altervista.org/VUIsProject/request-to-function-call---Chrome-STT.html (works in Chrome only)
- https://lucianoabriata.altervista.org/VUIsProject/request-to-function-call---Whisper-STT.html

And for a complete experience, the reader is referred to the example videos and screenshots shown in the next section.

## Section 3: Integration into our growing AR/VR molecular graphics tool

In the last months of work before submitting this report we were able to integrate a pipeline using Chrome's ASR coupled to GPT-4o-mini for transcription-to-function casting as shown above, into a draft prototype of our AR/VR molecular graphics web app. The app itself still requires a lot of work, but the VUI is quite complete thanks to the time we could dedicate to it with this grant.

As of the date of submission of this report, the VUI system is able to convert inputs spoken (or typed) in natural language into actions that enable control of the web app quite efficiently and seamlessly. Thanks to this, then, users can for example ask the system to "increase temperature by 50 degrees" to which the web app responds by increasing the temperature of the molecular simulation accordingly. Likewise, users can ask the system to start or stop the simulation, zoom in or out, change visualizations, etc.

Moreover, our LLM prompt optimized for GPT-4o-mini allows it to make "smart" unit conversions, cleanup of transcription errors, and translation from languages other than English (while the ASR module transcribes requests only from English, typing the requests works in any language). These positive aspects are not unexpected, as comparative studies show that LLM-based ASR can outperform general models in certain low-resource and code-switching settings, highlighting the importance of contextual language understanding.[9,10]

And as we append user requests and LLM outputs to a chat-like history, the system can keep track of a conversation; then, a user can for example say "color all atoms in red" followed by "no wait, you better color them by atom" and the system will understand both requests. This last specific feature is important for fluent and natural communication, and is accomplished by these kinds of lines in the input-output pairs of the LLM's prompt:

```
{ role: "user", content: "Increase temperature" },
{ role: "assistant", content: "changeTemperature(+30);" },
{ role: "user", content: "Again" },
{ role: "assistant", content: "changeTemperature(+30);" },
```

*Capabilities and limitations*

The best way to see this in action and thus get a grasp on the capabilities is of course through actual examples. Accordingly, we have recorded a video on a prototype integrating the ASR system developed here with our growing tool for molecular graphics, from inside a Meta Quest 3 device running our web app in AR mode (video demo available at https://www.youtube.com/watch?v=DozF99mW_KU and example screenshot in figure below). Note that since our ASR system of choice only runs in Chrome, we had to run this on a computer that manages the session (made possible by the multiuser nature of the tool).

The methodology comes of course with limitations, which we observed mainly on handling noisy environments and non-standard accents. This however happened to the two ASR systems tested, both of which performed best in quiet settings with no background noise (on top of which Whisper suffered from the problems described earlier).

Initial usability tests indicated positive reception from users, particularly regarding the intuitive nature of the interface. However, we need to test this much deeper, which will only be possible once we have a more complete app where users can run immersive interactive molecular graphics and modeling and actually require the VUI for real-world requests.

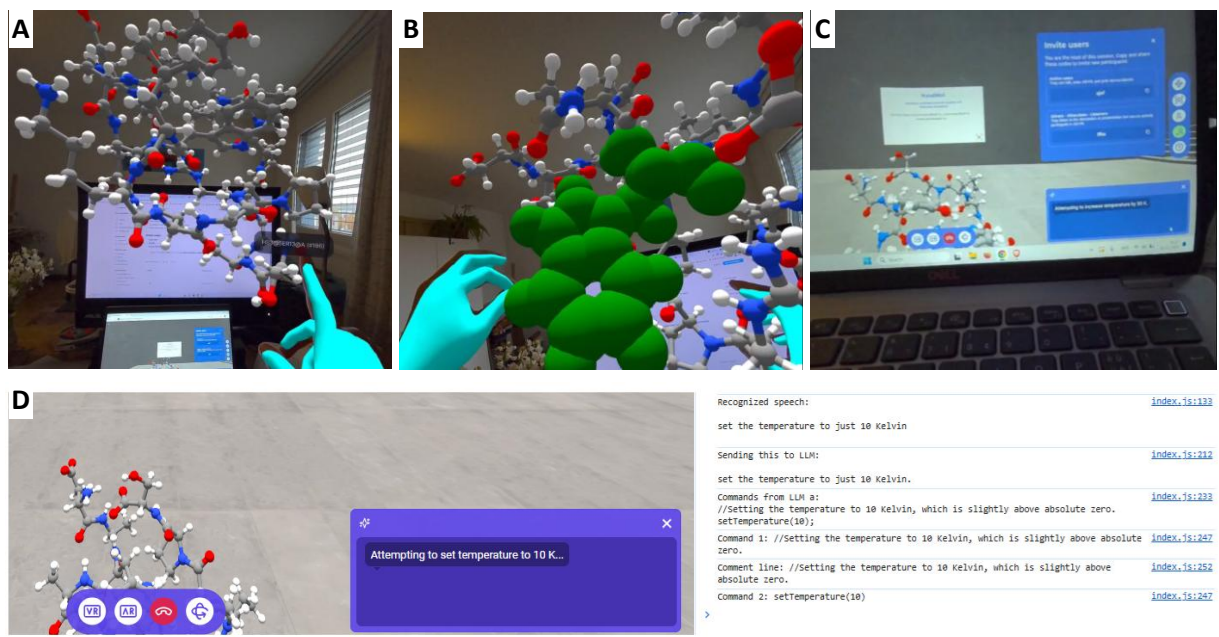

**Screenshots from the LLM-coupled, Chrome-based VUI as integrated in draft form into out molecular graphics app.** (A-C) Screenshots off the demo video shot from inside an AR/VR headset and available at https://www.youtube.com/watch?v=DozF99mW_KU. (A) and (B) show the basic immersive functionality and how users can interact with the molecular graphics to obtain or point at information (A) and with the MD simulation to act on it (B). In both cases, the screens in the back show the Host user's side, running in a Chrome browser to utilize its built-in speech recognition system. In (C) the view is focused on the AI/LLM VUI (blue box at the bottom right) which is executing a command after the user asked it to increase the simulation temperature. (D) Screenshot from the web browser where the Host is running speech recognition, with a focus on the VUI/AI module and a view of the console logs to show how speech is first transcribed and then transcripts are converted by the LLM into function calls and comments.

### Section 4: General and broader impact

This project provided us with a great deal of valuable experience in integrating ASR and NLP systems into web applications. Learning to optimize prompts for casting text into function calls was also very valuable, as it opened many possibilities for wider applications in domains beyond molecular graphics and modeling. Some such applications are already under the cook; for example, after seeing our ASR-NLP pipeline in action a group that develops software for genomics and transcriptomics data analysis approached us to see if we could help them develop a module that uses the same approach to allow users to process and analyze genomics and transcriptomics data through natural language requests. This relates to the most solid part of our results, that is going from text either typed or transcribed into function calls or code generation; therefore, we think we will be able to help this group in this exciting application—which would in turn illustrate the power of the methods proposed here in a field that has much more exposure given the relevance of those data analysis in fundamental biology and in medicine.

Finally, for others to benefit directly from this work you have funded, we have made all our results, code and prompts available online for free at the URLs provided throughout the report (but API keys are required). And once this report is approved, we will put it up as a blog post and/or preprint, under open access.

**Section 5: Challenges and future directions**

Several challenges emerged during this project, as already detailed throughout the report but briefly reviewed here. First, expanding Chrome's ASR grammar to improve recognition of domain-specific terms was not possible, limiting its accuracy for specialized commands. However, we could overcome some of these problems by instructing the downstream LLM molecule to detect and correct possible misspellings. See for example the section "*Note that some words might be wrong in the input provided due to errors upon speech recognition, for example 'some' might need to be replaced by 'zoom', 'so mean' could be 'zoom in', 'change' might actually be 'chain' or 'chains'*" in the system part of the prompt provided under "Final implementation" in section 2.

Whisper's tendency to hallucinate was a big disappointment, but it is possible that future AI-based ASR systems could be better than it and also better than the ASR system built into Chrome.

Similarly, the code-generation approach for ASR-to-action is tempting given its superior power; therefore, a more careful and dedicated study might be worthy—although several downsides were confirmed and even some new ones discovered, as detailed under *Approaches tested* in section 2.

One aspect we didn't touch upon in the report is the possibility that the web app replies back orally. While not originally in the proposed work, we did test briefly with Chrome's own speech synthesis API and also with a third-party API called Talkify who offered us free credits for testing. We found both to work very well, actually better than speech recognition by any system, so we started, by the end of the project, adding some speech synthesis for a more fluid interface in our program (some generated audio examples are featured in the video link provided to demonstrate the app under *Capabilities and limitations* in section 3).

Further work will focus on further optimizing the LLM prompt and addressing the system's robustness under different conditions. We will have to face this when we progress further with the app for molecular graphics and modeling, especially as we start testing it with actual users.

**Conclusion**

Overall, we call this project a success, because we observed that Chrome's ASR together with transcription cleanup by GPT-4o-mini upon conversion to function calls worked reasonably well, providing us with a clear way to implement flexible, natural language, speech-based commanding of web apps. As in any work well done, identifying the limitations was important, so that we properly inform users about what to expect and how to best interact with the tool in natural language.

We envision that if the implementation into the app for molecular graphics and modeling keeps going this well, the VUI will make the app unique, and at the same time this could foster future development of VUIs for controlling other software.